\documentclass[prb,twocolumn]{revtex4}% Physical Review B
\usepackage{epsfig}
\usepackage{graphicx}% Include figure files
\usepackage{bm}% bold math
\usepackage{amsfonts}

\usepackage[matrix,frame,arrow]{xy}
\usepackage{amsmath}

\newcommand{\ket}[1]{\left\vert{#1}\right\rangle}
\newcommand{\qw}[1][-1]{\ar @{-} [0,#1]}
\newcommand{\qwx}[1][-1]{\ar @{-} [#1,0]}
\newcommand{\gate}[1]{*{\xy *+<.6em>{#1};p\save+LU;+RU **\dir{-}\restore\save+RU;+RD **\dir{-}\restore\save+RD;+LD **\dir{-}\restore\POS+LD;+LU **\dir{-}\endxy} \qw}
\newcommand{\control}{*!<0em,.025em>-=-{\bullet}}
\newcommand{\ctrl}[1]{\control \qwx[#1] \qw}
\newcommand{\targ}{*!<0em,.019em>=<.79em,.68em>{\xy {<0em,0em>*{} \ar @{ - } +<.4em,0em> \ar @{ - } -<.4em,0em> \ar @{ - } +<0em,.36em> \ar @{ - } -<0em,.36em>},<0em,-.019em>*+<.8em>\frm{o}\endxy} \qw}
\newcommand{\multigate}[2]{*+<1em,.9em>{\hphantom{#2}} \qw \POS[0,0].[#1,0];p !C *{#2},p \save+LU;+RU **\dir{-}\restore\save+RU;+RD **\dir{-}\restore\save+RD;+LD **\dir{-}\restore\save+LD;+LU **\dir{-}\restore}
\newcommand{\ghost}[1]{*+<1em,.9em>{\hphantom{#1}} \qw}
\newcommand{\gategroup}[6]{\POS"#1,#2"."#3,#2"."#1,#4"."#3,#4"!C*+<#5>\frm{#6}}
\newcommand{\rstick}[1]{*!L!<-.5em,0em>=<0em>{#1}}
\newcommand{\lstick}[1]{*!R!<.5em,0em>=<0em>{#1}}
\newcommand{\Qcircuit}[1][0em]{\xymatrix @*[o] @*=<#1>}
\begin{document}

\title{Implementation of the three-qubit phase-flip error correction code with superconducting qubits}

\author{L. Tornberg$^1$, M. Wallquist$^2$, G. Johansson$^1$, V.S. Shumeiko$^1$, and G. Wendin$^1$}
\affiliation{$^1$ Chalmers University of Technology, SE-41296
Gothenburg, Sweden \\
%\author{M. Wallquist}
%\affiliation{ 
$^2$Institute for Theoretical Physics, University of Innsbruck, and Institute for Quantum Optics and Quantum Information of the Austrian Academy of Sciences,
\\ 6020 Innsbruck, Austria.}
%\author{G. Johansson}
%\author{V.S. Shumeiko}%
%\author{G. Wendin}
%\affiliation{ Chalmers University of Technology, SE-41296
%Gothenburg, Sweden.}

\date{\today}
\begin{abstract}
We investigate the performance of a three qubit error correcting
code in the framework of superconducting qubit implementations. Such a code can recover a quantum state perfectly in the case of dephasing errors but only in situations where the dephasing rate is low. Numerical studies in previous work have however shown that the code does increase the fidelity of the encoded state even in the presence of high error probability, during both storage and processing. In this work we give analytical expressions for the fidelity of such a code. We consider two specific schemes for qubit-qubit interaction realizable in superconducting systems; one $\sigma_z\sigma_z$-coupling and one cavity mediated coupling. With these realizations in mind, and considering errors during storing as well as processing, we calculate the maximum operation time allowed in order to still benefit from the code. We show that this limit can be reached with current technology. \\
\end{abstract}


\maketitle
\section{Introduction}

Recent experimental achievements in coupling
superconducting qubits \cite{Pashkin, Berkley, Steffen, Niskanen, Plantenberg, Majer,Sillanpaa} open up for investigating physical realizations of
simple few-qubit algorithms in such systems. One example of such a code is the three-qubit quantum error-correction
code (QECC) \cite{Preskill,Steane}, which is able to 
correct for a \emph{single} type of error acting on the physical qubits. This QECC has successfully been implemented in other systems \cite{Cory, Knill, Chiaverini,Boulant} proving the possibility to prolong the lifetime of a logical qubit by redundantly encoding its quantum state into three physical qubits. \\
In this paper we put the three qubit QECC into the context of quantum computation with superconducting qubits. The dominant noise in such systems is dephasing, that is the loss of phase coherence between the computational states in the system. This is  a process that can be modeled as a single type error\cite{NielsenChuang}, and can thus be corrected for by the QECC in question. However, the code can only recover the state perfectly under the assumption that errors on more than one qubit at a time can be neglected, and that the gates are instantaneous in time. If this is not the case, uncorrectable errors will occur during the storage of the quantum state as well as during the execution of the gates. In superconducting qubit systems, the dephasing rate is unfortunately comparable to the coupling energy between the qubits. In such systems the above assumptions are not justifiable and correctable errors will be accompanied by uncorrectable ones. In this paper we investigate the performance of the three qubit QECC for realistic gates in superconducting systems with relevant dephasing times. We address the question of how fast the two-qubit gate operation times need to be in order to benefit from coding, and if this limit can be reached with current technology. \\
Previous work addressing the questions of uncorrectable errors and noise during processing have either focused on continuous error correction where the gate operations are instantaneous in time \cite{Pellizzari,Paz,Ahn} or relied on numerical studies to investigate the effect of noise during the gate operations \cite{Chuang,Barenco}. In the regime of fast gate operations, the effects of correlated noise have also been studied \cite{Averin}. In this paper we take a different approach, where all \emph{single} qubit operations are considered to be much faster than  the dephasing time $t_{sqo}\ll T_2$. We can therefore neglect the errors that occur on this time-scale. As the two-qubit gate we consider the controlled-phase (cPhase) gate which is diagonal in the computational basis. Two realizations of this gate, relevant for superconducting implementations, are considered. In the first case, the coupling Hamiltonian itself is diagonal (experimentally achievable via e.g. a large Josephson junction \cite{Lantz}). This allows us to calculate the fidelity of the corrected state analytically. In the second case the qubit-qubit coupling is mediated by a cavity bus \cite{Wallquist}  where the dipole coupling between the cavity and qubits yields a Jaynes-Cummings type of interaction. Similar systems have recently attracted much attention due to their long coherence times \cite{Wallraff, Koch}. In this case the coupling is however not diagonal and we have to calculate the fidelity of the QECC numerically. We show that the fidelity of the QECC when using the cavity mediated coupling is comparable to the case when the diagonal cPhase gate is used.  \\
The structure of the paper is as follows. In section \ref{sec:instant} we briefly discuss the model used to describe dephasing and introduce the three qubit QECC with instantaneous gates. In section \ref{sec:direct} and \ref{sec:cavity} we consider realistic implementations of gates. The diagonal coupling is considered in section \ref{sec:direct} and the cavity mediated coupling in section \ref{sec:cavity}. 
We conclude in section \ref{sec:conclusion}.

\section{Instantaneous gates}\label{sec:instant}
We begin by describing the QECC in the ideal case, where only single errors are present, and the gates are instantaneous. In section \ref{sec:prolong} we relax the first approximation and study the case of multiple errors. \\
The Hamiltonian for a qubit coupled longitudinally to a heat bath of harmonic oscillators is given by 
\begin{equation}
H = -\frac{E}{2}\sigma_z + \sigma_z\sum_i\hbar\lambda_i(b_i + b_i^\dagger) + \sum_i\hbar\omega_ib_i^\dagger b_i
\end{equation}
where $E$ is the qubit level splitting and $b_i^\dagger/b_i$ the usual creation/annihilation operators of the harmonic oscillator. With this Hamiltonian, one can derive a master equation in the Markov limit \cite{Mahklin}
\begin{equation}\label{eq:dephDiff}
\dot{\rho}(t) = -\frac{\Gamma_\varphi}{2}(\rho(t) - \sigma_z\rho(t)\sigma_z),
\end{equation}
where $\Gamma_\varphi$ is the dephasing rate given by the zero frequency component of the noise spectral density 
\begin{equation}\label{eq:nsd}
L(\omega) = 2\pi\lambda^2(\omega)\eta(\omega)\coth\left(\frac{\hbar\omega}{2k_BT} \right),
\end{equation}
where $\hbar\lambda(\omega)$ is the coupling energy to the bath and $\eta(\omega)$ the bath density of states. We leave these quantities unspecified since the experimental parameter of interest is the dephasing rate itself. Solving Eqn. (\ref{eq:dephDiff}) gives an exponential decay of the off-diagonal elements in the density matrix 
\begin{equation}\label{}
\rho_{ij} =  e^{-\Gamma_\varphi t}\rho_{ij}(0). 
\end{equation}
The diagonal elements are stationary in time. The results for the diagonal and off-diagonal elements can be combined into the solution 
\begin{equation}\label{eq:rhoT}
\rho(t) = p(t)\rho(0) + \big(1-p(t)\big)\sigma_z \rho(0)\sigma_z 
\end{equation}
where $p(t)$ can be given the meaning of a time-dependent probability for the qubit to be in the correct state
\begin{equation}\label{eq:p}
p(t)= \frac{1}{2}[1 + \exp(-\Gamma_\varphi t)].
\end{equation}
In this paper we assume the dephasing to act independently on each qubit. In the presence of noise all three qubits in the circuit will thus evolve according to Eq.~(\ref{eq:rhoT}). 
%This solution can be recognized as an operator sum representation of the time evolution of a quantum system with the Krauss operators \cite{Preskill} 
%\begin{eqnarray}\label{eq:errDeph}
%E_0(t) &=& \sqrt{p(t)}\mathbf{1} \nonumber \\
%E_1(t) &=&\sqrt{1-p(t)}\sigma_z.
%\end{eqnarray}
This form allow us to interpret dephasing as a stochastic process where a phase flip occurs with probability $1-p(t)$. This discretization of the continuous phase damping is the very core of quantum error correction. 
\subsection{One qubit error}\label{sec:EC-describe}
\begin{figure}
\mbox{
\Qcircuit @C = 1.5em @R = 1em {
&  &\mbox{Encoding} && &&\mbox{Decoding} && & \\
 \lstick{\ket{\psi}} & \ctrl{1} & \ctrl{2}  & \gate{H} & \multigate{2}{\mathcal{E}_z(\rho)} & \gate{H} & \ctrl{2} & \ctrl{1} & \targ     & \rstick{\ket{\psi}} \qw\\
 \lstick{\ket{g}}    & \targ    & \qw       & \gate{H} &\ghost{\mathcal{E}_z(\rho)}        & \gate{H} & \qw      & \targ    & \ctrl{-1} &  \qw \\ 
\lstick{\ket{g}}     & \qw      & \targ     & \gate{H} &\ghost{\mathcal{E}_z(\rho)}        & \gate{H} & \targ    & \qw      & \ctrl{-1} &  \qw \gategroup{2}{2}{4}{3}{1em}{--}\gategroup{2}{7}{4}{8}{1em}{.} 
} 
}
\caption{The three qubit QECC for correcting phase-flip errors. The state of the information
carrying qubit
is encoded with
three qubits, using two ancilla qubits which are initially in the state
 $|g\rangle$. After decoding, a single phase-flip on the information
carrying qubit is corrected by a Toffoli gate, controlled by the two
ancilla qubits. A single qubit operation, conditioned by a measurement of the
two ancilla qubits, can substitute for the Toffoli gate. 
%The error $\mathcal{E}_z$ represent the phase-flip error.
}
\label{fig:idealProtocolPhaseflip}
\end{figure}
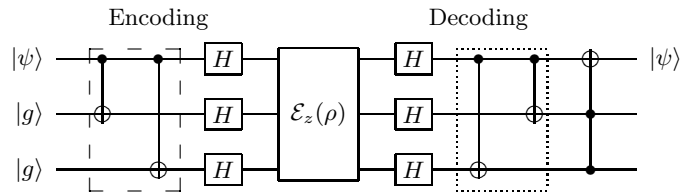

The three qubit QECC that corrects for phase flips is depicted in Fig. \ref{fig:idealProtocolPhaseflip}. 
The qubit is initially in the state $|\psi\rangle = \alpha|g\rangle + \beta|e\rangle$ and the ancillas are initiated in the ground state $|gg\rangle$. The first part of the circuit (dashed box) serves to entangle the three qubits leaving them in the encoded state
\begin{equation}\label{}
|\Psi\rangle = \alpha|ggg\rangle + \beta|eee\rangle.
\end{equation}
The role of the Hadamard gates is to rotate the phase flip of Eq. (\ref{eq:rhoT}) into a bit flip $H\sigma_z H = \sigma_x$. Thus, during the part of the evolution where errors occur, all three qubits will evolve according to Eq. (\ref{eq:rhoT}), but with $\sigma_z$ replaced with $\sigma_x$. If the dephasing is weak $\Gamma_\varphi t \ll 1$, the error probability is small $1-p(t) \ll 1$, and we can neglect terms representing errors occurring on more than one qubit at a time. Just before the decoding part (dotted box), the three qubits are in a mixed state with the following probabilities to find the system in the respective pure states
\begin{eqnarray}\label{eq:error}
p^3      &:& \alpha|ggg\rangle + \beta|eee\rangle \nonumber\\
p^2(1-p) &:& \alpha|egg\rangle + \beta|gee\rangle \nonumber\\
p^2(1-p) &:& \alpha|geg\rangle + \beta|ege\rangle \nonumber\\
p^2(1-p) &:& \alpha|gge\rangle + \beta|eeg\rangle. 
\end{eqnarray}
After the decoding the states are given by
\begin{eqnarray}\label{}
p^3      &:& (\alpha|g\rangle + \beta|e\rangle)|gg\rangle \nonumber\\
p^2(1-p) &:& (\alpha|e\rangle + \beta|g\rangle)|ee\rangle \nonumber\\
p^2(1-p) &:& (\alpha|g\rangle + \beta|e\rangle)|eg\rangle \nonumber\\
p^2(1-p) &:& (\alpha|g\rangle + \beta|e\rangle)|ge\rangle,
\end{eqnarray}
after which the Toffoli gate is applied, flipping the state of the first qubit iff the ancillas are in the state $|ee\rangle$. We see that this rotates the erroneous qubit state back into the correct one while leaving the others unaltered. The ancillas are then traced out leaving the qubit in the original pure state $|\psi\rangle$. Note that all states in Eq. (\ref{eq:error}) are mutually orthogonal, which is the reason why the code can detect and recover the state with perfect fidelity. More explicitly, the three qubits span an 8-dimensional Hilbert space, which can sustain 4 mutually 2-dimensional orthogonal subspaces. Each error resides in one subspace, and the code word in the remaining one. Due to the mutual orthogonality, the different errors can be distinguished and appropriate measures can be taken to rotate them back into the code space. From this it is clear that three different errors is all the three qubit QECC can correct for.\\
When the error probability is no longer small, one has to consider all the terms in the time evolution of the density matrix. Since this introduces additional errors it is clear that the QECC can no longer preserve the state with perfect fidelity. In the remaining part of the paper we consider this situation and investigate how the lifetime of the quantum state can be prolonged using realistic values for dephasing  and gate operation times in superconducting systems. 
%\begin{equation}
%p(1-p)^2 : \alpha|eeg\rangle + \beta|gge\rangle .
%\end{equation}  
%
\subsection{Multiple qubit errors}\label{sec:prolong}
\begin{figure}[!ht]
\includegraphics[width=8cm]{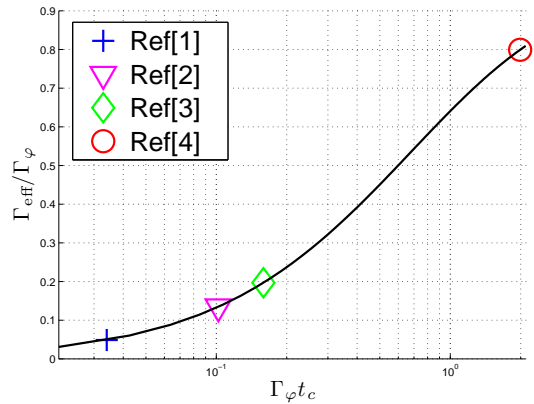}
\caption{(Color online) The ratio $\Gamma_{\mathrm{eff}}/\Gamma_\varphi$ plotted as a function of the renormalized correction time $\Gamma_\varphi t_c$. We mark four realistic values of $\Gamma_{\mathrm{eff}}/\Gamma_\varphi$ which can be achieved using current technology in superconducting qubits.}
\label{fig:effectiveDephRate}
\end{figure}

We begin by considering the ideal case where all gates are assumed to be implemented on a time scale much shorter than the dephasing time. We thus assume that no errors occur during the execution of the gates. To reach this limit in superconducting implementations more work  on reducing noise is however still required [REF]. Hence the results of this section will only be used as a benchmark for situations where this approximation is relaxed and gate errors are included. \\
From the time dependence of $p(t)$ the
error probability can no longer be considered
small as $\Gamma_\varphi t > 1$. As discussed in section \ref{sec:EC-describe}, we must take into account higher order terms and include processes where errors occur on two and three qubits simultaneously.  
%we write the time evolved three-qubit density matrix 
%\begin{equation}\label{eq:realErr}
%\rho(t) = \sum_{\alpha, \beta, \gamma} E_\alpha^{(1)} E_\beta^{(2)} E_\gamma^{(3)}
%\rho(0) E_\gamma^{\dagger(3)} E_\beta^{\dagger(2)},E_\alpha^{\dagger(1)},
%\end{equation}
%where the superscripts refer to the individual Hilbert spaces of the three qubits and $\alpha, \beta, \gamma \in\{ 0,1\}$, indexing the Krauss operators of Eq. (\ref{eq:errDeph}). 
The initial
density matrix, $\rho(0)$, is given by the pure three qubit state
used to encode the logical bit, $|\Psi\rangle = \alpha|000\rangle
+ \beta|111\rangle$. 
%The effect of dephasing is calculated by inserting the Krauss operators corresponding to dephasing, Eq. (\ref{eq:errDeph}), into Eq.
%(\ref{eq:realErr}). All physical qubits will thus be affected by the
%errors, 
After the occurrence of the errors we decode and correct according to the scheme
presented in Sec. \ref{sec:EC-describe}. The two ancilla qubits are
then traced out leaving a corrupted single qubit mixed final state
\begin{equation}\label{eq:rhoDephC}
\rho(t)=p_c(t)\rho(0) + (1-p_c(t))\sigma_x \rho(0)\sigma_x,
\end{equation}
with the probability \cite{Chuang}
\begin{equation}\label{eq:pc}
p_c(t) = \frac{1}{2}\left[1 + (3
- e^{-2\Gamma_\varphi t})e^{-\Gamma_\varphi t}/2\right].
\end{equation}
Comparing Eq. (\ref{eq:p}) and (\ref{eq:pc}) we see that if the correction procedure is applied often enough, $\Gamma_\varphi t \ll 1$, the probability to be in the correct state can be improved from  $\mathcal{O}(\Gamma_\varphi t)$ to $\mathcal{O}\big((\Gamma_\varphi t)^2\big)$ \cite{Chuang}. This regime is not accessible in all devices. We see however that $p(t) \leq p_c(t)$ for all $t$, making it clear that it always is beneficial to do the error correction in the above situation. With this in mind we now calculate how much the dephasing rate can be reduced with realistic parameters.
%\begin{figure}[!ht]
%\includegraphics[width=7cm]{fidelityMinDephasing.eps}
%\caption{The minimum fidelity $F^{\mathrm{min}}_c$ normalized to the minimum fidelity of the no coding case $F^{\mathrm{min}}$. The inset shows $F^{\mathrm{min}}_c$ and $F^{\mathrm{min}}$ for $\Gamma_\varphi t_c \ll1$.  }
%\label{fig:Fmin}
%\end{figure}
From Eq. (\ref{eq:rhoDephC}) we make the observation that the error and correction either flips the state of the qubit or leaves it unaffected. Repeating the procedure does not alter this situation. Thus in principle we can correct repeatedly and derive an effective dephasing rate. The probability, $p_{c,n}$ to be in the correct
state after $n$ such correction cycles obeys the equation
\begin{equation}\label{eq:repPcDeph_a}
\left(\begin{array}{c} p_{c,n} \\
1-p_{c,n} \end{array}\right)=\left( \begin{array}{cc}
p_c(t_c) & 1-p_c(t_c) \\
1-p_c(t_c) & p_c(t_c)
\end{array}\right)^n \left(\begin{array}{c} 1 \\
0 \end{array}\right),
\end{equation}
where $(1,0)$ and $(0,1)$ denote the correct and flipped state
respectively and $t_c$ is the duration of the cycle. This equation is easily solved yielding 
\begin{equation}\label{eq:repPcDeph}
p_{c,n}= \frac{1}{2}\left( 1 + \frac{(3 - e^{-2\Gamma_\varphi t_c})^n}{2^{n}}e^{-\Gamma_\varphi nt_c}\right).
\end{equation}
Comparing Eq. (\ref{eq:p}) and (\ref{eq:repPcDeph}) with $t =
nt_c$ we may derive an effective dephasing rate,
$\Gamma_{\mathrm{eff}}$, such that 
$p_\mathrm{eff}(t) = \frac{1}{2}\left[1 +
e^{-\Gamma_{\mathrm{eff}}t} \right]$.
The effective rate $\Gamma_{\mathrm{eff}}$ is related to $t_c$ and
$\Gamma_\varphi$ according to
\begin{eqnarray}
\Gamma_\mathrm{eff} &=& \Gamma_\varphi\left[ 1 - \frac{1}{\Gamma_\varphi
t_c}\ln \left(\frac{3-e^{-2\Gamma_\varphi t_c}}{2} \right) \right],
\end{eqnarray}
giving $\Gamma_\mathrm{eff}\leq \Gamma_\varphi$ for all $ \Gamma_\varphi
t_c$ as can be seen in Fig \ref{fig:effectiveDephRate} where the ratio $\Gamma_\mathrm{eff}/\Gamma_\varphi$ is plotted as a function of $\Gamma_\varphi t_c$. We also indicate the value of $\Gamma_{\mathrm{eff}}/\Gamma_\varphi$ that can be achieved with current examples of superconducting qubit implementations. Here we have estimated the repetition time as $t_c \sim \hbar/g$, $g$ being the qubit-qubit coupling energy.
We see that, even for realistic parameters, there can be a significant increase in coherence in the encoded quantum state. This shows that the three qubit QECC can be used to prolong the lifetime of the qubit even when the assumption of instantaneous gates is relaxed.
% 
%\section{Realistic gates}\label{sec:realistic}
%
\section{Direct qubit coupling}\label{sec:direct}
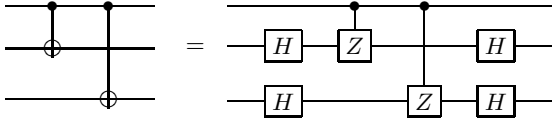
\begin{figure}
\mbox{
\Qcircuit @C = 1.5em @R = 1.4em {
  & \ctrl{1} & \ctrl{2}  & \qw\\
  & \targ    & \qw       & \qw\\ 
  & \qw      & \targ     & \qw 
} 
}
\mbox{ \Qcircuit @C = 1em @R = 1.7em {
&   &\\
& = &
}
}\mbox{
\Qcircuit @C = 1.5em @R = 1em {
& \qw      & \ctrl{1}        & \ctrl{2} &  \qw         & \qw\\
& \gate{H} & \gate{Z}\qw    & \qw      & \gate{H} \qw & \qw\\
& \gate{H} &    \qw          & \gate{Z} \qw      & \gate{H} \qw & \qw
}
}
\caption{ Gate sequence for encoding and decoding. The CNOT gates can be implemented using the diagonal cPhase gate ${\rm diag}[1,1,1,-1]$ together with single-qubit Hadamard gates. For this specific gate sequence, the Hadamard gates can be pulled out to the beginning and the end of the sequence.}
\label{fig:encoding}
\end{figure}

\begin{figure}
\mbox{
\Qcircuit @R = .7em @C = .5em {
 & \gate{\mathcal{E}(t_g)} &  \ctrl{1}    &\qw  & \ctrl{2}  & \gate{H} & \multigate{2}{\mathcal{E}(t_s)} & \gate{H} & \ctrl{2}  & \qw & \ctrl{1}  & \gate{\mathcal{E}(t_g)} &  \targ  & \qw\\
 & \gate{H}  & \gate{Z} \qw & \qw &\qw & \gate{\mathcal{E}(t_g)}  & \ghost{\mathcal{E}(t_g)}  & \gate{\mathcal{E}(t_g)} &  \qw    & \qw     & \gate{Z}\qw & \gate{H} & \ctrl{-1}& \qw\\
 & \gate{H} & \qw & \qw & \gate{Z} \qw & \gate{\mathcal{E}(t_g)}  & \ghost{\mathcal{E}(t_g)}  & \gate{\mathcal{E}(t_g)} & \gate{Z} \qw & \qw &\qw         & \gate{H}  & \ctrl{-1}& \qw
}
}
\caption{The equivalent circuit when errors occur during the
execution of the cPhase gates. The gate operation time and storage time are denoted $t_g$ and $t_s$ respectively. Since the error operators $\mathbb{I}$ and $\sigma_z$ now commute with all but the Hadamard gates, the error can be moved to the end or beginning of the cPhase gates. Hence the errors on the ancillas can be collected into a single one, simplifying calculations. } \label{fig:circuit2}
\end{figure}
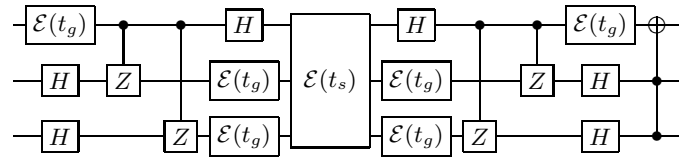

\begin{figure}[!ht]
\includegraphics[width=9cm]{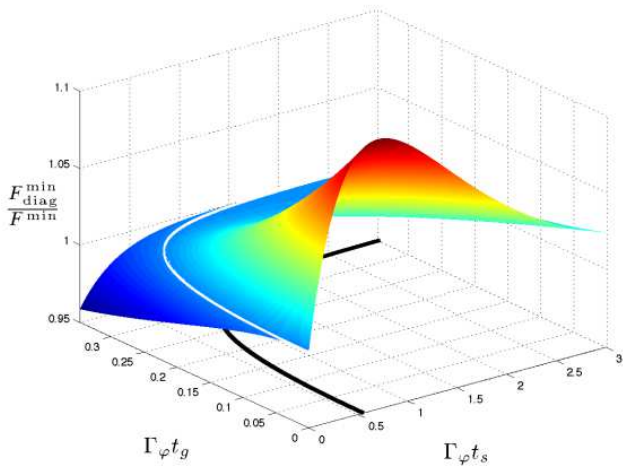}
\caption{(Color online) The normalized minimum fidelity $F_\mathrm{diag}^\mathrm{min}/F^\mathrm{min}$ for the case when errors occur during the gate operations. The region where $F_\mathrm{diag}^\mathrm{min} \geq F^\mathrm{min}$ is bounded by the white line. The black line shows the optimal storage time $t_s^\mathrm{opt}$.} \label{fig:areaEigen}
\end{figure}

We now depart from the approximation of perfect processing and consider the
regime where $\Gamma_\varphi t_{op} \lesssim1$, taking into account the errors that occur during the two-qubit gate operations. We however make the assumption of fast single qubit operations $t_{sqo} \ll 1/\Gamma_\varphi$ and neglect any dephasing occurring on this time-scale. We first consider the performance of the QECC protocol when the information-carrying qubit is
coupled directly to each ancilla qubit with a coupling which is
diagonal in the energy eigenbasis
\begin{equation}
H = \sum_{i}  E_{i} \sigma_z^{(i)} + \lambda_{i} \sigma_z^{(i)}\sigma_z^{(i+1)}.
\end{equation}
Such a coupling can be realized using e.g. the circulating currents in ring-shaped CPB qubits \cite{Vion} interacting via a large Josephson junction \cite{Lantz}. This coupling, together with single qubit phase gates naturally gives rise
to the general controlled-phase gate; ${\rm diag}[1,1,1,\exp[i4\lambda T/\hbar]]$, which for the interaction time
$\lambda T=\hbar\pi/4$ generates the cPhase gate. The advantage of studying the implementation of QECC with this setup first,
is 1) that the cPhase gate is also a natural gate for the cavity-mediated qubit coupling, and 2)
using the cPhase gate to implement the cNOT gates as shown in Fig. \ref{fig:encoding}, the error operators $\mathbb{I}$ and $\sigma_z$ (see Eq.~(\ref{eq:rhoT})) commute with all gates inside
the circuit except for the single-qubit Hadamard gates.
This implies that the errors occurring during the execution of the cPhase can be moved to the end or beginning of the gate as indicated in
Fig.\ref{fig:circuit2}. Denoting the gate operation time $t_g$ and the storage time $t_s$, we see that the ancillas are subject to the same errors as before, but now during a longer time $2t_g + t_s$. The error on the uppermost line, however, must be divided into three parts, since it is separated by the Hadamard gates. Calculating $\rho(t)$ thus reduces to the calculation of $2\cdot 8\cdot 2 = 32$ terms in the operator sum representation, which is analytically tractable (c.f. $8\cdot 8\cdot 8 = 512$ terms if the errors on the ancillas could not be collected). Since the Toffoli
gate can be replaced with fast measurements we also neglect errors
occurring during the execution of this. 
The final state of the qubit after detection and correction is given by the density matrix
\begin{eqnarray}\label{eq:rhofinalGates}
\rho(t) &=& p_1 \rho_1 +p_x \rho_x + p_y \rho_y +p_z \rho_z \nonumber \\
\rho_1& =& \quad1\rho1:  p_1 \equiv P_+\left[p(t_g)^2 +
(1-p(t_g))^2\right]\nonumber \\
\rho_z &=& \sigma_z\rho\sigma_z: p_z \equiv 2P_+ p(t_g)(1-p(t_g))\nonumber \\
\rho_x &=& \sigma_x\rho\sigma_x:  p_x \equiv P_-\left[p(t_g)^2 +
(1-p(t_g))^2\right] \nonumber \\
\rho_{y} &=& \sigma_y\rho\sigma_y:  p_y \equiv
2P_-p(t_g)(1-p(t_g)),
\end{eqnarray}
with $p(t_g)$ as in Eq.~(\ref{eq:p}) and the probabilities $P_+$ and $P_-$ given by
\begin{equation}\label{eq:PQ}
P_\pm = \frac{1}{4}\left( 2 \pm e^{-\Gamma t_s} \pm 2e^{-\Gamma( 2 t_g + t_s)} \mp e^{-\Gamma(
4 t_g + 3 t_s)} \right).
\end{equation}
Since there now is more than one type of error present in the final state of the qubit, it is no longer meaningful to compare the error probabilities directly. Instead we use the fidelity \cite{NielsenChuang} 
\begin{equation}\label{eq:fidDef}
F(t) = \langle \psi|\rho(t)|\psi \rangle,
\end{equation}
between the initial state $|\psi\rangle$ and the corrupted state $\rho(t)$ of Eq. (\ref{eq:rhofinalGates}) to quantify the benefit of using the QECC. Since $F(t)$ depends on the initial state $|\psi\rangle$ we use the minimum fidelity as a measure of the code performance
\begin{equation}\label{eq:minFid}
F^\mathrm{min} = \min_{|\psi\rangle} F(t).
\end{equation}
To find this minimum fidelity, we make the observation that $P_+ \geq P_-$ for all times $t_g, t_s$. This, together with the fact that $p(t_g)^2 + (1-p(t_g))^2\geq 2p(t_g)(1-p(t_g))$ for all $t_g$ gives us the inequalities (see Eq.~(\ref{eq:rhofinalGates})) $p_1 \geq p_z \geq p_y$ and $p_x \geq p_y$ 
%\begin{equation}\label{}
%\min_i p_i  = p_y,\qquad i = 1,x,y,z,
%\end{equation}
This in turn gives the minimum fidelity according to
\begin{eqnarray}\label{eq:Fmin_ge}
%F_\mathrm{diag} =  \sum_i p_i(t) \langle\psi|\rho_i|\psi\rangle,
F_\mathrm{diag} &=&  p_1 + p_x\langle \sigma_x\rangle^2+ p_y\langle \sigma_y\rangle^2+ p_z\langle \sigma_z\rangle^2 \geq \nonumber \\
&& p_1 + p_y (\langle \sigma_x\rangle^2 + \langle \sigma_y\rangle^2 +\langle \sigma_z\rangle^2) = \nonumber \\
&& p_1 + p_y \equiv F_\mathrm{diag}^\mathrm{min}.
\end{eqnarray} 
%$F(t)$ depends on the initial state $|\psi\rangle$ we use the minimum fidelity as a measure of the code performance
%\begin{equation}\label{eq:minFid}
%F^\mathrm{min} = \min_{|\psi\rangle} F(t).
%\end{equation}
%
%\begin{equation}\label{eq:fidDephNoCodeMin}
%F^\mathrm{min} = p(t),
%\end{equation}
Thus, we find the minimum fidelity for a $\sigma_y$-eigenstate, as opposed to Ref.~\onlinecite{Chuang} where the input state that minimizes the fidelity lies in the $xz$-plane between $|0\rangle$ and $(|0\rangle + |1\rangle)/\sqrt{2}$, depending on the ratio $t_s/t_g$. \\
We now compare the minimum fidelity between the state in Eq. (\ref{eq:rhofinalGates}) and the uncorrected state 
which is subject to the same dephasing given by
\begin{equation}\label{eq:fidDephNoCodeMin}
F^\mathrm{min} = p(t),
\end{equation}
with $t = 2t_g + t_s$. The normalized minimum fidelity $F_\mathrm{diag}^\mathrm{min}/F^\mathrm{min}$ is plotted in Fig. \ref{fig:areaEigen}. For perfect gates, coding improved the fidelity for all times. The situation is drastically different when errors occur during processing. There is now a lower limit on the speed of the gate operations, given by $F^\mathrm{min} = F_\mathrm{diag}^\mathrm{min}$, above which QECC is beneficial. This condition can equally be stated as a relation between $t_g$ and $t_s$ given by 
\begin{equation}\label{eq:areaEigen}
t_s =
\frac{1}{2\Gamma}\left( -6\Gamma_\varphi t_g - \log\left[ e^{-4\Gamma_\varphi
t_g}(2-e^{2\Gamma_\varphi t_g})\right] \right),
\end{equation}
which is plotted as the white line in Fig. \ref{fig:areaEigen}. 
If we believe that error correction is crucial to the realization of large scale quantum computing this limit sets the standard for how fast processing needs to be. The maximum time $t_g^\mathrm{max}$ allowed in order to benefit from the error correction is easily obtained from Eq. (\ref{eq:areaEigen})
\begin{equation}\label{eq:t1Max}
t_g^\mathrm{max} = \frac{\log 2}{2 \Gamma_\varphi}.
\end{equation}
For typical values of the dephasing time in superconducting systems, $t_\varphi \simeq 1 \mu s$, this gives a maximum gate operation time of the order $0.1 \mu s$.  
From Fig. \ref{fig:areaEigen} it is clear that, for each value of $\Gamma_\varphi t_g$ there exists an optimum storage time such that the difference $F_\mathrm{diag}^\mathrm{min} - F^\mathrm{min}$ is maximized
\begin{equation}\label{eq:t_sOptimum}
t_s^\mathrm{opt} = -\frac{1}{2\Gamma_\varphi}\left[ 2\Gamma_\varphi t_g + \log\left( \frac{2 - e^{2\Gamma_\varphi t_g}}{3}\right)\right].
\end{equation}
This is plotted with a solid black line in the $t_st_g$-plane of Fig. \ref{fig:areaEigen}. \\
We can understand the origin of the optimum in the following way. For instantaneous gates $t_g = 0$, we recover the result of section \ref{sec:prolong} where coding is beneficial for all finite storage times $t_s$. Along the line of zero storage time $t_s = 0$ however, we see there is nothing to be gained from using the code. This is simply because the logical qubit is only shielded when it is encoded, i.e. during the storage. For zero storage time, we are only applying a series of faulty gates, which of course reduces the fidelity. As $t_s$ grows, there is a competition between these uncorrectable errors and the beneficial effects of the code obtained for any finite $t_s$. As $t_g$ exceeds the limit given in Eq. (\ref{eq:t1Max}) the positive effects of the code can however no longer compensate for the uncorrectable errors in the gates. Finally, as $t_s \to \infty$, there is no longer any advantage in using the code and the fidelity approaches that of the unprotected qubit.\\
We now consider what happens with the condition given in Eq. (\ref{eq:t1Max}) when the code is repeated $n$ times, each run taking time $t_c^{ge} = 2t_g + t_s$. The four
probabilities, $p_i^n$, describing the state of the qubit after
$n$ such cycles obey the equation
\begin{equation}\label{eq:PnGates}
\left(
\begin{array}{c}
p_{1,n} \\
p_{z,n} \\
p_{x,n} \\
p_{y,n}
\end{array}\right) = \underbrace{
\left(
\begin{array}{cccc}
p_1 & p_z & p_x & p_y \\
p_z & p_1 & p_y & p_x \\
p_x & p_y & p_1 & p_z \\
p_y & p_x & p_z & p_1 \\
\end{array}\right)^n}_{\equiv P_\mathrm{ge}^n} \left(
\begin{array}{c}
1 \\
0 \\
0 \\
0
\end{array}\right),
\end{equation}
in the basis $[\rho,\rho_z,\rho_x,\rho_y]^T$. It is useful to express the probabilities $p_i^n$ in term of the eigenvalues of the $P_\mathrm{ge}$-matrix 
\begin{eqnarray}\label{eq:pnLambda}
p_{1,n} &=& \frac{1}{4}(1 + \lambda_2^n)(1 + \lambda_3^n) \nonumber \\
p_{z,n} &=& \frac{1}{4}(1 + \lambda_2^n)(1 - \lambda_3^n) \nonumber \\
p_{x,n} &=& \frac{1}{4}(1 - \lambda_2^n)(1 + \lambda_3^n) \nonumber \\
p_{y,n} &=& \frac{1}{4}(1 - \lambda_2^n)(1 - \lambda_3^n),
\end{eqnarray}
where the eigenvalues are given by 
\begin{eqnarray}\label{}
\lambda_1 &=& 1 \nonumber\\
\lambda_2 &=& 1 - 2(p_x + p_y) \nonumber\\
\lambda_3 &=& 1 - 2(p_z + p_y) \nonumber\\
\lambda_4 &=& \lambda_2\lambda_3 .
\end{eqnarray}
From Eq. (\ref{eq:pnLambda}) and the fact that $\lambda_i \geq 0$ for all times, it is clear that $\min_i p_{i,n} = p_{y,n}$. The minimum fidelity in Eq. (\ref{eq:Fmin_ge}) thus generalizes to the iterative case by replacing $p_1$ with $p_{1,n}$ and $p_y$ with $p_{y,n}$.  Further the condition for the QECC to be beneficial, $F_\mathrm{diag}^\mathrm{min} \geq F^\mathrm{min}$, in the iterative case gets the form
\begin{eqnarray}\label{eq:areaEigen_n}
p_{1,n} + p_{y,n}  \geq p(nt_c^{ge}) 
\end{eqnarray}
which equivalently can be expressed as
\begin{eqnarray}\label{eq:compareEigenvalues}
\lambda_3 \lambda_2 \geq \tau_2,
\end{eqnarray}
where $\tau_2 = 2p(t_c^{ge}) -  1$ is the eigenvalue $\neq 1$ of the corresponding $P$-matrix for the case of no error correction. Since Eq. (\ref{eq:compareEigenvalues}) is $n$-independent it holds for any $n$. In particular we can choose $n = 1$ for which it coincides with the condition in Eq. (\ref{eq:t1Max}). Since the code introduces additional errors this result is not obvious. We had rather expected a case where a combination of the different probabilities in Eq. (\ref{eq:PnGates}) would result in a faster fidelity decay.  \\
The main conclusion from this section is that the three qubit QECC can improve the fidelity of the qubit state, even in the case of errors during gate operations. In order to benefit from the code, one must however assure that the normalized gate operation time $\Gamma_\varphi t_g$ can be made shorter than $ \log 2/2 \approx 0.35$. We emphasize that the results in this section hold for a diagonal qubit coupling only.
%We deliberately chose this coupling due to its simple form, making it possible to analytically study the QECC. With the result of this simple coupling in mind, we are now in a position to evaluate the performance of the cavity mediated qubit-qubit coupling. This is the topic of the next section. 
\section{Coupling via tunable cavity}\label{sec:cavity}
\begin{figure}[!ht]
\includegraphics[width=8cm]{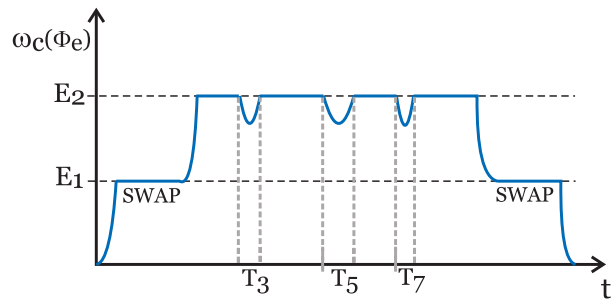}
\caption{(Color online) Pulse sequence for performing a control-phase gate by tuning the frequency of the relevant cavity
mode. The sequence begins and ends with a "swap" - a single photon $\pi$-pulse while the cavity is on resonance with qubit 1. During the resonance with qubit 2, single-photon $\pi$-pulses and two-photon $\pi$-pulses are mixed; the idle periods $T_3$ and $T_7$ in between are chosen to annihilate excited photon states, and $T_5$ is chosen to create the non-trivial phase shift.
}
\label{fig:cPhaseprot}
\end{figure}

\begin{figure}[!ht]
\includegraphics[width=8cm]{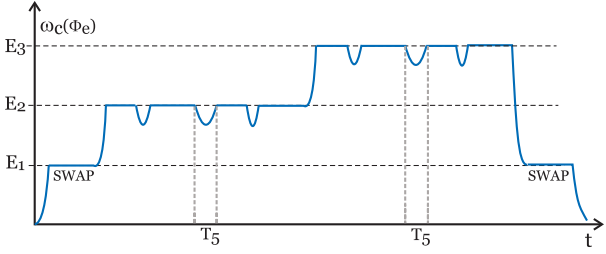}
\caption{(Color online) Pulse sequence for performing the two cPhase gates involved in the encoding/decoding protocol. As described in the text, one can move the cavity mode directly from resonance with qubit 2 to qubit 3 without performing the intermediate $\pi$-pulses in resonance with qubit 1, thus making the sequence shorter and easier to implement.}
\label{fig:encod_prot}
\end{figure}

The system we consider for implementing the QECC is a set of Cooper-pair box (CPB) qubits \cite{Shnirman} which are capacitively connected
to a one-dimensional superconducting stripline cavity. The strong-coupling regime has been achieved experimentally for
this type of mesoscopic cavity-QED system \cite{Wallraff}. 
In this setup, the stripline is terminated with a SQUID whose effective inductance can be tuned by applying an external magnetic flux $\Phi_e$. Thus by changing $\Phi_e$ one changes the cavity boundary conditions which leads to a tunable resonance frequency $\omega_c(\Phi_e)$ of the cavity \cite{Wallquist}. The Hamiltonian of the decoupled cavity mode and the CPB qubits reads, at the charge degeneracy point, in the qubit eigenbasis
\begin{equation}\label{eq:H0}
H_0 = \hbar \omega_c (\Phi_e) \left(a^\dagger a + {1 \over 2}\right) - \sum_{i=1}^3 {E_{J,i} \over 2} \sigma_z^{(i)},
\end{equation}
where $E_{J,i}$ is the Josephson energy of the $i$th qubit. In the idle state, the cavity mode is assumed to be far off resonance with all the qubits, with weak dispersive cavity-qubit
coupling which in the following is considered negligible. Further, to allow the cavity to address the qubits separately, we assume the qubits
to be separated in frequency space, with $|E_{J,i} - E_{J,j}|$ larger than the qubit-cavity coupling. 
Tuning the cavity mode into resonance with a specific qubit, $\omega_c (\Phi_e) = E_{J,i}$, the interaction term obtains the
familiar Jaynes-Cummings form,
\begin{equation}\label{eq:Hint}
H_{int,i} = i{\hbar g_i \over 2} \left(
a^\dagger \sigma_-^{(i)} - a \sigma_+^{(i)}\right).
\end{equation}
In Ref.~[\onlinecite{Wallquist}] it has been shown how to perform the universal two-qubit cPhase gate using the cavity as a bus mediating the qubit interaction. The strategy is to tune the oscillator through resonance with one qubit at a time, performing $\pi$-pulse swaps in every step, and then reverse the sequence. However, due to the photon number dependence of the
$\pi$-pulse swap operation time, $T= \pi/g\sqrt{n+1}$, 
one has to insert interference loops in the
interaction between qubit 2 and the oscillator \cite{Schmidt-Kaler}, interrupted by idle periods $T_n\ll 1/g$ where qubit and
cavity are decoupled as shown in figure \ref{fig:cPhaseprot}.
Note that the oscillator must be tuned adiabatically
not to excite higher oscillator states, $\dot\omega_c/\omega_c \ll\omega_c$.\\
In the implementation of the QECC-protocol in Fig. \ref{fig:idealProtocolPhaseflip}, 
two consecutive cNOT operations are used. The gate sequence for the encoding is equivalent to 
two consecutive cPhase gates in addition to
single qubit Hadamard gates, as shown in figure \ref{fig:encoding}. Let us focus on the operation of the cavity and forget single qubit gates for the moment since these are
performed independently of the cavity. In the first step of the cPhase operation, the state of qubit 1 is swapped onto the oscillator,
$(\alpha |g\rangle + \beta |e\rangle)|0\rangle \rightarrow |g\rangle (\alpha |0\rangle + \beta |1\rangle)$. During the
next steps the oscillator interacts with qubit 2 in order to create a phase shift which depends on {\it both} the state of the oscillator 
{\it and} 
the state of qubit 2. The last step is to swap the oscillator state back onto qubit 1:
$|g\rangle (\alpha |0\rangle + \beta |1\rangle) \rightarrow (\alpha |g\rangle - \beta |e\rangle)|0\rangle$ (the minus sign
appears because of the form of the interaction term). When we perform two cPhase gates after each other, first involving qubits 1 and 2, then
involving qubits 1 and 3, it is possible to shorten the sequence. 
Namely, at the end of the first cPhase gate the state of the oscillator is transferred to qubit 1, but then it is 
immediately transferred back to the oscillator at the beginning of the second cPhase sequence. 
 At the very end, these two operations, $|g\rangle (\alpha |0\rangle + \beta |1\rangle) \rightarrow (\alpha |g\rangle - \beta |e\rangle)|0\rangle
\rightarrow |g\rangle (\alpha |0\rangle - \beta |1\rangle)$, correspond only to a single qubit
rotation on qubit 1. Thus we can shorten the encoding protocol by directly tuning the oscillator from qubit 2 to qubit 3, as shown in
figure \ref{fig:encod_prot}, and correcting the sign with an additional diagonal single-qubit gate.\\
%Having described the implementation of universal qubit gates in the cavity-qubit system, we now move on to investigate the performance of the QECC in the non-optimal case of multiple errors.    
%
%%%%%%%%%%%%%%%%%%%%%%%%%%%%%%%%%%%%%%%
%
\subsection{Simulating the 3 qubit + cavity system with dephasing}
During the pulse sequence described in the previous section, the qubits will interact with each other by entangling their states with the oscillator. If dephasing is added to this picture it is no longer tractable to solve the equation of motion analytically to obtain a solution as that in Eqn. (\ref{eq:rhoT}). Instead we simulate the system dynamics by solving its governing master equation numerically. This section is divided into two parts. In the first part we describe how to derive master equation from a system plus bath interaction. In the second part we discuss how this equation is implemented numerically. The major difficulty with the analysis is that the system Hamiltonian will depend on whether the qubit is on or off resonance with the cavity. Because of this, we must be careful when treating the coupling to the bath, which will have different structure in the energy eigenbasis of the two different cases. Another difference lies in the fact that we have introduced an additional quantum system trough the cavity. This is also subject to dephasing which must be taken into account. The standard procedure is to couple the cavity to a heat bath longitudinally through the cavity number operator \cite{Gardiner}. Since only one qubit at a time is on resonance with the cavity, our analysis can, without loss of generality, be limited to the case of a single qubit plus cavity. The Hamiltonian for the system plus baths will then be given by 
\begin{eqnarray}\label{eq:HFull}
H = H_{sys} + \sigma_zX + nY + H_{bath}^X + H_{bath}^Y, 
\end{eqnarray}
where the baths are chosen to be a collection of harmonic oscillators and $n= a^\dagger  a$ is the cavity number operator. In this case $X = \hbar\sum_i\lambda_i(b_i + b_i^\dagger)$ and $H_{bath}^{X} = \sum_i\hbar\omega_i b_i^\dagger b_i$ with the oscillator bath defined analogously. The system Hamiltonian $H_{sys}$ is given by 
\begin{equation}
H_{sys} = \left\{
\begin{array}{l}
H_0 = \hbar \omega_c  \left(a^\dagger a + {1 \over 2}\right) - {E_{J} \over 2} \sigma_z ,\qquad \mathrm{off~resonance}\\
H_{JC} = H_0 + i{ \hbar g \over 2} \left(
a^\dagger \sigma_- - a \sigma_+\right),\quad \mathrm{on~resonance}\\
\end{array}\right.
\end{equation}
for the idle and resonant periods in the pulse sequence respectively. 
 
\subsubsection{Dephasing with qubit and cavity on resonance}\label{sec:dressedDeph}
In this section we show how to derive a master equation for the reduced density matrix of the qubit plus cavity system. For simplicity, we only consider the qubit-bath coupling. The cavity can be treated analogously and we refer to Appendix \ref{app:A} for a detailed analysis. In the idle periods the system Hamiltonian commutes with the qubit-bath interaction. The situation is thus identical to that described by Eqn. (\ref{eq:dephDiff}) which means that the coupling to the bath results in pure dephasing in the energy eigenbasis $|g/e,n\rangle$
\begin{equation}
\rho_{g,n;e,m}(t) = \delta_{mn}e^{-\Gamma_\varphi t}\rho_{g,n;e,m}(0).
\end{equation}
When the qubit is moved into resonance with the cavity the system is described by the JC-Hamiltonian whose eigenstates are the dressed states with corresponding eigenenergies 
%\begin{equation}\label{eq:dressedstates}
%\begin{array}{ll}
%|n;\pm\rangle = \frac{1}{\sqrt{2}}(|g;n\rangle \pm i|e;n-1\rangle), & |g;0\rangle \\  E_{n;\pm} = \hbar\omega n \mp \frac{g\sqrt{n}}{2} & E_{g;0} = 0
%\end{array}
%\end{equation}
\begin{equation}\label{eq:dressedstates}
\begin{array}{ll}
|n;\pm\rangle = \frac{1}{\sqrt{2}}(|g;n\rangle \pm i|e;n-1\rangle), & E_{n;\pm} = \hbar\omega n \mp \frac{\hbar g\sqrt{n}}{2}, \\ |g;0\rangle,  & E_{g;0} = 0.
\end{array}
\end{equation}
What is important from the point of view of dephasing is that in the dressed state picture, the longitudinal coupling to the bath becomes transversal with matrix elements
\begin{equation}
\begin{array}{ll}\label{eq:matrixelememts}
\langle 0;g|\sigma_z|0;g\rangle = 1,   & \langle m;\pm|\sigma_z|n;\pm\rangle = 0, \\
\langle 0;g|\sigma_z|n;\pm\rangle = 0, & \langle m;\pm|\sigma_z|n;\mp\rangle = \delta_{mn}.
\end{array}
\end{equation}
Hence in this basis the coupling will not induce pure dephasing, but will instead give rise to relaxation between the dressed states. We note that this operator is block-diagonal which means that the relaxation will take place within the  blocks of equal $n$. Within such blocks the system behaves approximately like a two-level system coupled transversally to a bath with interaction Hamiltonian $H = \sigma_xX$. Hence, the diagonal elements of the density matrix will approach equilibrium exponentially with the relaxation rate $1/T_1$ given by the noise spectral density $L(\omega)$  (see Eqn. (\ref{eq:nsd})), 
%\begin{equation}\label{eq:rate}
%\frac{1}{T_1} = 2\pi\lambda^2(\Omega)\eta(\Omega)\coth\left(\frac%{\hbar\Omega}{2k_BT} \right),
%\end{equation}
at the relevant transition frequencies given by $\omega =g\sqrt{n}$ as can be seen from Eqns. (\ref{eq:dressedstates}) and (\ref{eq:matrixelememts}).
Similarly, coherences formed by superpositions of states within blocks of equal $n$ will decay exponentially with a rate $1/T_2 = 1/2T_1$. One important difference between the quasi two-level systems formed by the dressed state blocks and a real two-level system is that the relaxation rates for different blocks will be different due to the non-harmonicity of the dressed state energy levels. In addition to this, coherences formed by superpositions of states between different blocks will have another set of decay rates. (This can be seen from the full Liouville equation for the reduced density matrix, see Eqn. (\ref{eq:LiouvilleEqn}) and (\ref{eq:Liouvillian})). However, we note that all relevant transition frequencies lie in a range given by the cavity-qubit coupling $g$. Thus if we assume that the baths have no structure on this scale and that the temperature is much higher than the cavity-qubit coupling $k_BT\gg \hbar g$, we can safely approximate the noise spectral density in Eqn. (\ref{eq:nsd}) by its zero frequency limit for all relevant transitions. The situation is similar for the oscillator-bath coupling, for which we make the same assumptions about the bath. Apart from setting all the rates equal this approximation has another important implication. Since all rates are taken at zero frequency, they will be the same for the resonant and off-resonant passages. With this clearly stated we can derive a master equation in the Markov approximation for the reduced cavity~+~qubit density matrix
\begin{eqnarray}\label{eq:master}
\dot{\rho}  &=& -\frac{i}{\hbar}[H_{sys},\rho] - \gamma (n^2\rho + \rho n^2 - 2n\rho n) \nonumber \\
&-&\frac{\kappa}{2}(\rho - \sigma_z\rho\sigma_z),
\end{eqnarray}   
with the rates $\kappa$ and $\gamma$ given in Appendix \ref{app:A} along with a detailed derivation including the treatment of the oscillator bath. We emphasize that $\kappa$ and $\gamma$ are basis independent within our approximation.

\subsubsection{Numerical approach}
To treat Eq.~(\ref{eq:master}) numerically we project it on the instantaneous eigenbasis $|i\rangle$ of $H_{sys}$ to get it on Redfield form \cite{Rau} 
\begin{eqnarray}\label{eq:Bloch-Redfield}
\dot{\rho}_{ij} &=& -i\omega_{ij}\rho_{ij} - \sum_{kl}R_{ijkl}\rho_{kl},
\end{eqnarray}
where $\omega_{ij}\equiv (E_i - E_j)/\hbar$ is the energy difference between the eigenstates $|i\rangle$ and $|j\rangle$ and $\hat{R}$ denotes the Redfield tensor
\begin{eqnarray}\label{eq:Rtensor}
&R_{ijkl} & = \frac{\kappa}{2} \sum_{\mu=1}^3 \Big[\delta_{ik}\delta_{jl}
- \langle i|\sigma_z^{(\mu)}|k\rangle \langle
l|\sigma_z^{(i)}|j\rangle\Big] \\
& - & \gamma\Big[ \langle i|n^2|k\rangle\delta_{lj}+ 
\delta_{ik}\langle l|n^2|j\rangle - 
2\langle i|n|k\rangle\langle l|n|j\rangle\Big]\nonumber , 
\end{eqnarray}
where $\delta$ is the Kronecker delta and the summation $\mu$ runs over the number of qubits. Numerically, it is convenient to work in the Liouvillian space where $\rho$ is a vector quantity. In this way, we rewrite Eq. (\ref{eq:Bloch-Redfield}) as a matrix equation 
\begin{equation}\label{eq:matrixMaster}
\dot{\rho}_{[ij]} + i\hat{\omega}_{[ij][ij]}\rho_{[ij]}= \sum_{[kl]}\hat{R}_{[ij][kl]}\rho_{[kl]}. 
\end{equation}
The three qubits and oscillator span an $8(N + 1)$-dimensional Hilbert space, where $N$ is the number of photons in the resonator. This makes $\rho$ a column vector of length $64(N + 1)^2$. The Redfield tensor is a $64(N+1)^2\times64(N+1)^2$ matrix and $\omega$ is a diagonal matrix of the same size. We work in the eigenbasis of the qubit~+~oscillator Hamiltonian which we obtain by exact numerical diagonalization. The solution to Eq. (\ref{eq:matrixMaster}) is given by  
\begin{equation}\label{eq:matrixMasterSolution}
\rho(t) = e^{-(i\hat{\omega} + \hat{R})t}\rho(0) \equiv \hat{U}(t)\rho(0), 
\end{equation}
which can be solved by numerically diagonalizing the propagator $\hat{U}(t)$. In this way we sequentially simulate the entire pulse sequence with the output density matrix of one passage serving as the input of the next. In the spirit of the previous sections all errors occurring during single qubit operations are neglected. Hence, we are not concerned with the dynamics of these gates, which consequently are realized as matrix multiplications.  As in the previous sections we use the  minimum fidelity as the measure of code performance, which is obtained by numerically searching the space of initial states.  
\subsubsection{Numerical results}
\begin{figure}[!ht]
\includegraphics[width=9cm]{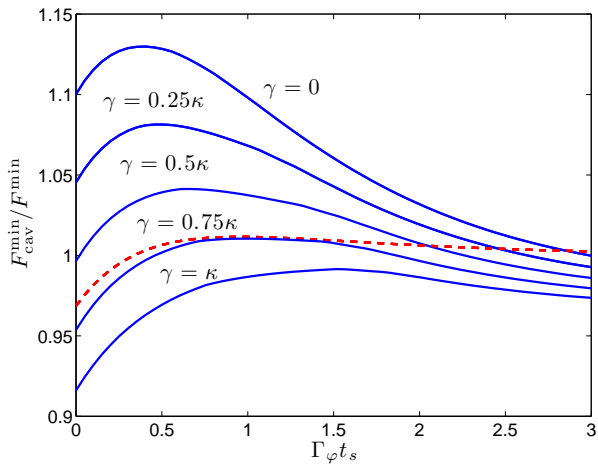}
\caption{(Color online) The ratio $F^\mathrm{min}_\mathrm{cav} / F^{\mathrm{min}}$ (solid lines) for different values of oscillator-environment coupling $\gamma$. The ratio $F^\mathrm{min}_\mathrm{diag} / F^{\mathrm{min}}$ is plotted (dashed line) for comparison. Using the minimum fidelity as measure, the two gates are comparable for $\gamma \simeq 0.75\kappa$.}
\label{fig:numericalFid}
\end{figure}

The goal of this section is to show that, for realistic values of qubit-oscillator coupling, the cavity mediated cPhase gate outperforms the diagonal coupling in the limit  $\gamma \to 0$ when the the oscillator is not damped. This is very encouraging for future applications, since resonators with lifetimes several orders of magnitude longer than qubits can be made \cite{Martin,Palacios}.\\  
The gate operation time $t_g$ is now set by the qubit oscillator coupling and the level splitting of the qubits and oscillator respectively. From experiments on similar systems we expect the coupling energy to be in the range  $g/h \simeq 10-100$ MHz \cite{Majer} for all qubits. To get restrictive results we chose $g/h = 18$ MHz as a typical value. This corresponds to a temperature $T\sim 0.8$mK, and is thus consistent with the high temperature approximation discussed in section \ref{sec:dressedDeph}. The qubit dephasing rate was set to $\Gamma_\varphi = 1$ MHz in all simulations. The qubit energies where taken to be $E_{J,i}/h = 4850, 5000$ and $5150$ MHz respectively. The energy separation was chosen to match the performance of current state of the art tunable oscillators in superconducting systems \cite{Martin,Palacios}. The oscillator frequencies in the idle periods were chosen to be $\omega_c/2\pi = 4925$ and 5075 MHz respectively. With these parameters the gate operation time is given by $\Gamma_\varphi t_g = 0.26$,\cite{Wallquist} which is in the range where coding improves the fidelity for the diagonal coupling. \\
We plot the minimum fidelity normalized to the minimum no-coding fidelity for several values of the cavity dephasing rate $\gamma$ in Fig. \ref{fig:numericalFid}. To compare the two implementations, the result for the diagonal gate (Eq. (\ref{eq:Fmin_ge})) is plotted in the same figure (dotted line).
When cavity and qubits are subject to the same amount of dephasing ($\gamma = \kappa$), the cavity mediated coupling cannot match the diagonal. This is however to be expected since we have introduced an additional uncorrectable channel of noise through the cavity. It is however interesting to see that the fidelity can be improved with a modest reduction in cavity-environment coupling. With a choice of $\gamma = 0.75\kappa$ the two couplings exhibit comparable fidelities as can be seen in Fig. \ref{fig:numericalFid}. Further decrease in $\gamma$ improves the situation even further. If storage times up to $\Gamma_\varphi t_s \simeq 2$ are considered, the cavity mediated coupling actually performs better than the diagonal coupling. We attribute this improvement to the SWAP-operations, which transfer the state from the qubit to the cavity. Given that the cavity is more phase coherent than the qubit the coherence is thus partially protected during part of the processing. 
%%%%%%%%%%%%%%%%%%%%%%%%%%%%%%%%%%%
\section{Conclusion}\label{sec:conclusion}
In this paper we have studied the performance of the three qubit phase-flip QECC for realistic gates in superconducting systems. Since such a quantum code requires a limited amount of qubits for its implementation, it is interesting from the point of view of superconducting devices, where such an experiment should be realizable in the near future. We have studied two explicit couplings, one diagonal and one cavity mediated. \\
Our analysis begin with the case of ideal gates, where no gate errors occur during processing. In this case, we show that coding is beneficial, not only in the short time limit, but can also significantly reduce the dephasing rate when considering realistic experimental parameters. We move on to consider realistic gates, deriving analytical expressions for the fidelity in the case where the qubits are coupled in diagonal fashion. We found an upper limit on the gate operation time $t_g$ allowed to benefit from coding. For the cavity mediated coupling, we study the system numerically, solving the master equation that describes dephasing in the qubits and cavity. For realistic values of cavity-qubit coupling we find that the fidelity is comparable to that of the much simpler diagonal coupling. In the limit of weak coupling between the environment and cavity, this coupling even outperforms the diagonal one. 
%We attribute this to the transfer of coherence between qubit and cavity, where it is protected during part of the pulse sequence. Since stripline cavities with high $Q$-values have been demonstrated, this is promising for future applications.
We attribute this effect to the transfer of coherence from the qubit to the cavity, where it is protected during part of the pulse sequence. In view of the high Q-values demonstrated for stripline cavities [REF], this is promising for future applications.
\section{Acknowledgments}\label{sec:ack}
This work is supported by the European Commission through the
IST-015708 EuroSQIP integrated project and by the Swedish Research
Council.
\appendix
\section{Derivation of the master equation.}\label{app:A}
The cavity plus qubit system with dephasing is modeled by coupling its constituents longitudinally to a thermal bath so that the Hamiltonian for the full system plus bath is given by Eqn. (\ref{eq:HFull}). In the Markov approximation, it is possible to derive an equation of motion for the reduced cavity plus qubit density matrix. To second order in the couplings  $\lambda_i$ this equation reads
\begin{eqnarray}\label{eq:LiouvilleEqn}
\dot{\rho}_{ij} &=& -i\omega_{ij}\rho_{ij} + \mathcal{L}_{ij}^{n,Y}(\rho)+\mathcal{L}_{ij}^{\sigma_z,X}(\rho)
\end{eqnarray}
where $\omega_{ij}\equiv (E_i - E_j)/\hbar$ is the energy difference between the eigenstates $|i\rangle$ and $|j\rangle$ of $H_{sys}$. The dissipative dynamics is governed by $\mathcal{L}(\rho)$ whose matrix elements are given by 
\begin{eqnarray}\label{eq:Liouvillian}
\mathcal{L}_{ij}^{o,Q}(\rho) &=& \sum_{kl} 
o_{ik}\rho_{kl}o_{lj} (L_Q^\ast(\omega_{jl}) +  L_Q(\omega_{ik})) \\ 
&-& o_{ik}o_{kl}\rho_{lj} L_Q(\omega_{kl}) - \rho_{ik}o_{kl}o_{lj}L_Q^\ast(\omega_{lk})\nonumber,
\end{eqnarray}
where $o$ is the system operator that couples to the bath and $L(\Omega)$ is the Laplace transform of the bath correlator
\begin{equation}
L_Q(\Omega) = \int_0^\infty d\tau e^{-i\Omega\tau} \langle Q(\tau)Q(0)\rangle.
\end{equation}
For the case of a bath in thermal equilibrium, the real part of $L_Q$, which determines the transition rates will be given by 
\begin{eqnarray}
\mathrm{Re}~L(\Omega) &=& \pi \eta\lambda^2(\Omega)\bar{n}(\Omega,T)\times \nonumber \\
&&\left( \exp\left(  \frac{\hbar\Omega}{k_BT} \right)\Theta(\Omega) + \Theta(-\Omega) \right),
\end{eqnarray}
where $\lambda(\omega)$ is the coupling energy to the bath, $\eta(\omega)$ the oscillator density of states, $\bar{n} = (\exp(\hbar\omega/k_BT)-1)^{-1}$ the Bose occupation number and $\Theta(x)$ is the step function where we adopt the convention $\Theta(0) = 1/2$. The imaginary part of $L_Q$ is responsible for a small energy shift (Lamb shift) of the energy levels. Hence it will not be relevant for any pure phase damping, which is the central interest from the point of view of error correction. This effect will therefore be disregarded in the following analysis. The relevant transition frequencies $\omega_{ij}$ will be determined by the matrix elements of $\sigma_z$ and $n$ respectively. We must therefore separate our analysis into two cases; the resonant and the off-resonant periods in the pulse sequence described in the previous section. In the off-resonant case, the eigenstates of the system are the product states $|g/e,n\rangle$ for which both $n$ and $\sigma_z$ are diagonal. This leaves us with only one relevant transition frequency $\omega_{ij} = 0$. When the qubit and cavity are on resonance, the eigenstates of the system are the dressed states of the Jaynes-Cummings Hamiltonian 
\begin{equation}
|n;\pm\rangle = \frac{1}{\sqrt{2}}(|g;n\rangle \pm i|e;n-1\rangle),
\end{equation}
for which the matrix elements of $n$ and $\sigma_z$ reads
\begin{eqnarray}
\langle m;\pm|\sigma_z|n;\pm\rangle &=& 0 \nonumber \\
\langle m;\pm|\sigma_z|n;\mp\rangle &=& \delta_{mn} \nonumber \\
%\end{eqnarray}
%and 
%\begin{eqnarray}
\langle m;\pm|n|n;\pm\rangle &=& \left(n-\frac{1}{2}\right)\delta_{mn} \nonumber \\
\langle m;\pm|n|n;\mp\rangle &=& \frac{1}{2}\delta_{mn}.
\end{eqnarray}
In this case we get three relevant frequencies $\omega_{n\pm,n\mp} = \pm g\sqrt{n}$ and $\omega_{n\pm,n\pm} = 0$. We now assume that the baths have no structure on the scale of the qubit-cavity coupling. We further assume the temperature to be much higher than the cavity-qubit coupling $k_BT\gg\hbar g$. In this case we may safely approximate $\mathrm{Re}L_Q(\pm g\sqrt{n}) \simeq \mathrm{Re}L_Q(0)$ and conclude that the only relevant parameter for dissipation will be $L_Q(0)$, for both the resonant and off-resonant regime. With this clearly stated we can, from Eqn. (\ref{eq:LiouvilleEqn}), write down the master equation for the system dynamics as given in Eqn. (\ref{eq:master}) with the rates $\kappa$ and $\gamma$ given by $\kappa = 4\mathrm{Re}L_X(0)$ and $\gamma = \mathrm{Re}L_Y(0)$ independently of the choice of $H_{sys}$. These are the dephasing rates for the bare qubit and cavity systems respectively.

%%%%%%%%%%%%%%%%%%%%%%%%%%%%%%%

%%%%%%%%%%%%%%%%%%%%%%%%%%%%%%%%%%%%%%%
\end{document}